\documentclass[12pt]{iopart}
\usepackage{graphicx}
\usepackage{color}
\usepackage{afterpage}

\begin{document}

\title{Overlapping modularity at the critical point of $k$-clique percolation}

\author{B. T{\'o}th $^1$, T. Vicsek $^1$ and G. Palla $^2$}

\address{$^1$Dept. of Biological Physics, E{\"o}tv{\"o}s Univ., 1117 Budapest, P{\'a}zm{\'a}ny P. stny. 1/A}
\address{$^2$Statistical and Biological Physics Research Group of HAS, 1117 Budapest, P{\'a}zm{\'a}ny P. stny. 1/A}

\eads{\mailto{btoth@hal.elte.hu}, \mailto{vicsek@hal.elte.hu}, \mailto{pallag@hal.elte.hu}}

\date{7th of December, 2012}

\begin{abstract}
One of the most remarkable social phenomena is the formation of communities
in social networks corresponding to families, friendship circles, work teams, 
etc. Since people usually belong to several different communities 
at the same time, the induced overlaps result in an extremely 
complicated web of the communities themselves. Thus, uncovering the intricate 
community structure of social networks is a non-trivial task with great
potential for practical applications, gaining a notable interest in 
the recent years. The Clique Percolation Method (CPM) is one of the
earliest overlapping community finding methods, 
 which was already used in the analysis of several different social networks.
In this approach the communities correspond to $k$-clique percolation clusters,
and the general heuristic for setting the parameters of the 
method is to tune the system 
just below the critical point of 
$k$-clique percolation. 
However, this rule is based on simple physical principles and its
validity was never subject to quantitative analysis.
Here we examine the quality of the partitioning
in the vicinity of the critical point using recently introduced
overlapping modularity measures. 
According to our results on 
real social- and other networks, the overlapping modularities show a maximum 
close to the critical point, justifying the original criteria for
the optimal parameter settings. 

\end{abstract}

\maketitle

\section{Introduction}
\label{sect:intro}

A widely used tool for the study of social phenomena is provided by 
networks, based on the fundamental concept of mapping the connections
among people into a graph. Due to the developments in Information Technology,
our social activities and relations generate various forms of data on
 the large scale. On the one hand this offers a gold mine for research,
and on the other hand it can cause non-trivial data handling problems. 
The network approach has turned out to be very successful in the study of 
large scale social data, marked by investigations on mobile-phone 
networks \cite{Kertesz_mobile,Kertesz_mobile_2,Blondel_mobile,Seshadri_mobile}, e-mail networks \cite{Bornholdt_email,Newman_email,De_Wilde_email,Watts_email}, co-authorship networks \cite{Tamas_coauth,Newman_coauth,Leyersdorf_coauth,Ramasco_coauth,Borner_coauth,Wagner_coauth} and online social 
networks \cite{Mislove_online,Nazir_online,Andrew_online,Menzer_folk_soc,Scifanella_folk_soc}. We note that the idea of representing a complex system with 
a network is frequently used in various other fields as well, 
including biology, computer science, economy, etc. 
According to a very
recent survey \cite{coll_motion}, the network approach can be useful also 
in the description of the collective motion of dynamically interacting agents.
A highly interesting
 feature of real networks is that in spite of their independent origin,
 they show many universal features,  characterised by  a low average distance
combined with a high average clustering coefficient,
anomalous degree distributions, spreading phenomena 
and correlations \cite{Watts-Strogatz,Laci_science,Laci_revmod,Dorog_book,barrat}. 

One of the most widely studied area of complex network research is 
devoted to communities, 
(also called as modules, clusters, cohesive groups, etc.), 
associated with
more highly interconnected parts
\cite{GN-pnas,Huberman,spektral,potts,CPM_nature,Infomap,Blondel,Ronhovde,Santo_modszer,Sune_nature}, (for a detailed review on communities see 
Ref.\cite{Fortunato_report}).
Such building blocks (functionally related  proteins \cite{ravasz-science,spirin-pnas}, industrial sectors \cite{onnela-taxonomy}, interconnected Autonomous
Systems in the Internet \cite{Sneppen_internet_coms}, similar blogs on the World Wide Web \cite{blog_comm_1,blog_comm_2}, etc.) can play a crucial role in forming the structural and functional
properties of the involved networks. Another field of growing interest
in complex network theory is related to hierarchy \cite{Tamas_fract,Havlin_fract,Newman_hier,Avetisov,our_PNAS,our_rotated}, and 
the presence of
 communities in networks is one of the relevant and informative signature 
of the hierarchical nature of complex systems
\cite{ravasz-science,vicsek-nature,self-sim-coms}. 

Communities play a central role in social network research as well, where they can correspond to families, friendship circles, professional teams, or on 
a larger scale to fan clubs, institutions, etc. \cite{scott-book,watts-dodds}. 
These different types
of modules show non-trivial behaviour from several aspects. E.g., the 
time evolution of smaller collaborative or friendship circles
shows significant differences when compared to larger communities like 
institutions \cite{our_com_evolv}. 
Another surprising result is the dissasortativity
of the graph of communities in highschool friendship networks \cite{Gonzalez_high_school}, especially
in the light of the assortative nature of social networks in general. 
Very large scale social communities are also highly interesting, e.g., 
in a study concerning the mobile phone network of Belgian users the arising
communities contained only adjacent municipalities, and the only community
running across the ``linguistic border'' between the Walloon- and Flemish
regions was the one related to the Brussels and surroundings \cite{Blondel_Belgian}.

Motivated by the importance of community finding in social networks,
 (and in complex networks in general), here we focus on a theoretical
problem related to the quality of partitioning. First of all,
 we point out that in case of social networks, allowing overlaps
between the communities is crucial, as we are all members of our family,
friendship circle, working group, etc., at the same time in parallel. 
(Several results suggest that overlaps between communities are important
also in biology, where e.g.,  proteins can be part of more than one functional
unit \cite{protein-complexes}). One of the first algorithms allowing
shared members between the communities was given by the clique percolation
method (CPM). Here the basic building blocks of the communities are given
by $k$-cliques, (complete sub-graphs of $k$-nodes), and
communities are associated with $k$-clique percolation clusters. 
The usual rule for finding the optimal partitioning in this approach is
to tune the system to the critical point of $k$-clique percolation. 
(In case of e.g., weighted networks this can be achieved by applying 
an appropriate weight threshold). 
The reasoning behind this rule is based on simple ``physical'' principles: 
the emergence of a giant percolating community  would merge 
(and make invisible) many smaller communities, thus, to find a community 
structure as highly structured as possible, one has to be at the critical
point, where the rising of a giant $k$-clique percolation cluster is
just avoided. 

Although the above approach for setting the parameters 
of the CPM was successful in producing meaningful communities in many 
real networks, the quality of the partitioning obtained 
this way was never compared to the results for different parameter settings. 
Thus, in this paper we examine the quality of 
the communities in the vicinity of the critical point of $k$-clique 
percolation. For quantifying the quality of the
community partitions, we rely on various recently introduced modularity
 measures \cite{Nepusz_fuzzy,Chen_version,Shen_modul,Nicosia_modul,Sune_nature,Dani_modul}, all designed specially for overlapping communities. 
 (The concept of maximising a real valued modularity function for finding 
the best community partition is a very popular approach in general, and
 is also used in non-overlaping clustering problems.)
The motivation of our 
 research is the following:
Since the idea of tuning the $k$-clique percolation to the critical point and modularity maximisation are two independent principles aimed at the same target, (i.e., finding optimal community partitions), it is an interesting question 
whether they show any
consistency with each other? I.e., if the partitioning is optimal at
the critical point also from the point of modularity, we should observe
a maximum in the modularity. We note however that this maximum should 
be treated 
as a ``local'' maximum, or more precisely as a maximum amongst the 
CPM partitions obtained at different stages of the $k$-clique
 percolation transition. 
The global maximum for the modularity may correspond to a partition
(amongst all possible community
partitions) which contains communities that are not $k$-clique percolation
clusters.


The paper is organised as follows. In
 Sect.\ref{sect:CPM}. we describe the CPM in short, 
while in Sect.\ref{sect:modul}. we overview the different overlapping 
modularity measures. In Sect.\ref{sect:Appl}. we examine the behaviour 
of the listed modularities in the vicinity of the critical point of 
$k$-clique percolation in real networks, and finally we conclude 
in Sect.\ref{sect:Concl}.

\section{The Clique Percolation Method}
\label{sect:CPM}
As mentioned in Sect.\ref{sect:intro}., the community 
definition in this approach is based on $k$-cliques. A $k$-clique is a sub-graph 
with maximal possible link density, (i.e., every member of a $k$-clique is 
connected to every other member), therefore,
 it is a good starting point for defining communities. However, a method
accepting only complete sub-graphs as communities would be too restrictive.
Therefore, $k$-cliques are ``loosen up'' in the following way. 
Two $k$-cliques are considered adjacent if they share $k-1$ nodes, and a 
community is defined as the union of $k$-cliques that can be reached 
from each other
through a series of adjacent $k$-cliques. In other words, a community
is equivalent to a $k$-clique percolation cluster. We note that a 
$k$-clique percolation cluster is very much like a regular edge
percolation cluster in the $k$-clique adjacency graph, where the nodes represent the
$k$-cliques of the original network, and there is a link between two
nodes, if the corresponding two $k$-cliques are adjacent. 
The two main advantages of the community definition above is its 
local nature that it allows overlaps between the communities: 
a node can be
 part of several $k$-clique percolation clusters at the same time.

When applied to weighted networks the CPM method can have two parameters: 
the $k$-clique size $k$, 
 and a weight threshold $w^*$ (links weaker than $w^*$ are ignored).
When $k$ and $w^*$ are very high, only a few disintegrated community remain,
 while for low $k$ and $w^*$ in many cases we see a giant community arising,
 spreading over the majority of the network. When varying the 
weight threshold at a fixed $k$, the transition from the dispersed
communities to the giant community is analogous to a percolation 
phase transition. (E.g., previous work has shown that the $k$-clique
percolation transition in the Erd\H{o}s-R{\'e}nyi graph is a generalisation
of the regular edge percolation transition \cite{our_prl,our_jsp}). 
The criterion for finding 
the optimal value of $w^*$  is based on the aim to find 
a community structure as highly as possible: when the threshold is 
high we neglect too many links (and communities), while
the giant community appearing at low $w^*$ values can smear out the details by
merging (and making invisible) the smaller communities. Thus, former
works suggested adjusting $w^*$ close to the critical point of $k$-clique
percolation, where we take into account as many links as possible without 
allowing the emergence of a giant community.

\section{Modularity measures}
\label{sect:modul}

\subsection{The modularity by Girvan and Newman}
The most popular quality function for community partitions is
given by the modularity of Newman and Girvan \cite{Newman_modul}, comparing 
the fraction of links inside the communities to the expected
fraction of links in a random graph where the individual node degrees 
are equal to the node degrees in the original network. The basic idea 
behind this approach is that the number of links inside well defined
communities should be significantly larger than what we would except
at random. The random null model serving as the reference point is
given by the configuration model \cite{Molloy_Reed}, 
where the probability for having a connection between nodes 
$i$ and $j$ with degrees $d_i$ and $d_j$ is given by $d_id_j/4M^2$, where
$M$ denotes the total number of links.
Accordingly, the expected fraction of links inside community $\alpha$ 
is expressed by first summing up the node degrees in $\alpha$ as $d_{\alpha}=\sum_{i \in \alpha}d_i$, and simply writing $(d_{\alpha}/2M)^2$. Based on the above, the modularity can be given as
\begin{equation}
Q=\sum_{\alpha=1}^KQ_{\alpha}=\sum_{\alpha=1}^{K}\left[\frac{l_{\alpha}}{M}-\left(\frac{d_{\alpha}}{2M}\right)^2\right],
\label{eq:Newman_modul}
\end{equation}
where $l_{\alpha}$ denotes the number of links inside community $\alpha$, 
(and $l_{\alpha}/M$ is simply the fraction of links in $\alpha$). 
We note that (\ref{eq:Newman_modul}) can be
also rewritten in a form with summation over the individual nodes as
\begin{equation}
Q=\frac{1}{2M}\sum_{ij}\left( A_{ij}-\frac{d_id_j}{2M}\right)\delta(\alpha_i,\alpha_j),
\label{eq:Newman_mod_alt}
\end{equation}
where $A_{ij}$ stands for the adjacency matrix, ($A_{ij}=1$ if $i$ and $j$ are
 linked, otherwise $A_{ij}$ is zero), and $\delta(\alpha_i,\alpha_j)$
ensures the exclusion of terms where $i$ and $j$ are in different communities.

\subsection{Fuzzy modularity} 
The original modularity (\ref{eq:Newman_modul}-\ref{eq:Newman_mod_alt}) 
defined for ``crisp'' partitions can be generalised for overlapping 
communities in different ways. A straight forward solution was proposed
by Nepusz et al.\cite{Nepusz_fuzzy} defining a fuzzy partition matrix 
$u_{\alpha i}$ in the following way. The column $i$ of
$u_{\alpha i}$ is listing how the membership of node $i$ is divided amongst 
the communities,
\begin{eqnarray}
& &0\leq u_{\alpha i}\leq 1, \\
& & \sum_{\alpha=1}^Ku_{\alpha i}=1,
\end{eqnarray}
i.e.,  $u_{\alpha i}=0$ when it is not a member at all in
community $\alpha$, whereas a non-zero $u_{\alpha i}$ signs a belonging
to $\alpha$ in some extent. In the limiting
case of a crisp partition all entries except one in the column become zero,
and the entry corresponding to the sole community of the node becomes one.
In parallel, the row $\alpha$ of $u_{\alpha i}$ is listing the membership
values of the nodes in community $\alpha$. The row sum $\sum_{i=1}^Nu_{\alpha i}$
can be treated as the generalisation of the community size. 

By introducing a scalar product between the column vectors of $u_{\alpha i}$ we
obtain a similarity measure between the nodes defined as
\begin{equation}
s_{ij}=\sum_{\alpha =1}^Ku_{\alpha i}u_{\alpha j},
\label{eq:similarity}
\end{equation}
where the summation is running over the communities. When nodes $i$ and $j$ 
have non-zero memberships in absolutely different communities, $s_{ij}=0$,
 while larger $s_{ij}$ values usually 
indicate more similar memberships vectors. In the limiting case of 
crisp partitions $s_{ij}=1$ if and only if $i$ and $j$ belong to the same community, or in other words, $s_{ij}$ becomes equivalent to 
$\delta(\alpha_i,\alpha_j)$. This observation leads naturally to the 
idea of replacing $\delta(\alpha_i,\alpha_j)$ by $s_{ij}$ 
in (\ref{eq:Newman_mod_alt}) for gaining an overlapping modularity
 measure as
\begin{equation}
Q_f=\frac{1}{2M}\sum_{ij}\left( A_{ij}-\frac{d_id_j}{2M}\right)s_{ij}.
\label{eq:fuzzy_modul}
\end{equation}
A very nice feature of the fuzzy modularity obtained in this way is that
in case of crisp partitions it is equivalent to the original modularity
by Newman given in (\ref{eq:Newman_modul}-\ref{eq:Newman_mod_alt}). 

However, not all overlapping community detection algorithms evaluate 
$u_{\alpha i}$ explicitly, instead they provide only the list of members
in each community. In this case several possibilities open up for
calculating $u_{\alpha i}$. The simplest idea is to divide the membership 
values of the nodes equally amongst their communities independently of the
underlying network topology \cite{Shen_Physica_A} as
\begin{equation}
u_{\alpha i}=\frac{1}{q_i},
\label{eq:part_matr_simp}
\end{equation}
where $q_i$ denotes the number of communities $i$ participates in. To take into
account the number of links between the community members and the communities  
Chen et al.\cite{Chen_version} instead proposed
\begin{equation}
u_{\alpha i} = \frac{\sum_{j \in \alpha} A_{ij}}{\sum_{\alpha'}\sum_{j \in \alpha'} A_{ij}}.
\label{eq:part_matr_Chen}
\end{equation}
 An even more sophisticated approach is suggested
by Shen et al.\cite{Shen_modul}, considering the maximal cliques
in the network and summing over all neighbours of a given member $i$ inside 
the community $\alpha$ as
\begin{equation}
u_{\alpha i}=\frac{1}{u_i}\sum_{j\in \alpha}\frac{C^{\alpha}_{ij}}{C_{ij}}A_{ij},
\label{eq:part_matr_Shen}
\end{equation}
where $C^{\alpha}_{ij}$ denotes the number of maximal cliques in $\alpha$ 
containing the link $(i,j)$, and $C_{ij}$ stands for the total number of maximal
cliques in the network containing the link $(i,j)$. The pre-factor $1/u_{i}$ 
is for normalisation and can be calculated as
\begin{equation}
u_i=\sum_{\alpha=1}^K\sum_{j\in \alpha}\frac{C^{\alpha}_{ij}}{C_{ij}}A_{ij}.
\end{equation}
The most general formulation of the fuzzy modularity (\ref{eq:fuzzy_modul})
was given by Nicosia et al.\cite{Nicosia_modul}, where the concept of 
comparing the observed number of links between community members to expectation
values based on random null-models was extended to overlapping communities.

\subsection{Alternative ideas for modularity}

Instead of generalising the terms in the original modularity 
(\ref{eq:Newman_modul}-\ref{eq:Newman_mod_alt}), another option for 
constructing a measure for the quality of overlapping
partitions is to build up a formula based on ``first principles'', 
i.e., combining terms expressing various criteria for 
well behaving communities with possible overlaps. 
Here we overview two different approaches
along this line.

\subsubsection{Partition density by Ahn et al.}

A very interesting approach for revealing overlapping 
communities was suggested by Ahn et al. \cite{Sune_nature}, 
based on clustering the links 
instead of the nodes. 
In this approach link pairs sharing a node are ordered according to the similarity
between the neighbourhoods of their other end points. By using a single-linkage
hierarchical clustering based on this similarity, we obtain a link dendrogram,
and cutting this dendrogram at some threshold yields overlapping communities
for the nodes. For determining the optimal cut, Ahn et al. 
defined the partition density for an individual community $\alpha$ as
\begin{equation}
D_{\alpha}=\frac{M_{\alpha}-(N_{\alpha}-1)}{N_{\alpha}(N_{\alpha}-1)/2-(N_{\alpha}-1)},
\end{equation}
where $N_{\alpha}$ and $M_{\alpha}$ denote the number of nodes and links inside
$\alpha$ respectively. Assuming that $\alpha$ is connected, 
$0\leq D_{\alpha}\leq 1$, i.e., when $\alpha$ is tree-like, $D_{\alpha}=0$, 
whereas a fully connected community receives $D_{\alpha}=1$. The partition
density $D$ for the whole system is given by the average of $D_{\alpha}$,
weighted by the fraction of links inside the communities \cite{Sune_nature}:
\begin{equation}
D=\sum_{\alpha=1}^K\frac{M_{\alpha}}{M}D_{\alpha}=\frac{2}{M}\sum_{\alpha=1}^K
M_{\alpha}\frac{M_{\alpha}-(N_{\alpha}-1)}{(N_{\alpha}-2)(N_{\alpha}-1)}.
\label{eq:partition_dens}
\end{equation}
A very nice feature of (\ref{eq:partition_dens}) compared to 
e.g., (\ref{eq:Newman_modul}-\ref{eq:Newman_mod_alt}) is its local nature,
preventing the emergence of the resolution limit observed in case
of the original modularity \cite{Fortunato_resolution}.

\subsubsection{The overlapping modularity by L\'az\'ar et al.}

Another alternative for the  overlapping modularity was proposed 
by L\'az\'ar et al.\cite{Dani_modul}. The first criterion for 
obtaining a well defined community in this approach is 
that the members should devote
the majority of their links to the community rather than other parts of
 the network. To quantify this aspect,
the contribution of member $i$ to the modularity of its community $\alpha$
is calculated by comparing the number of its neighbours inside $\alpha$ to the 
number of its neighbours outside $\alpha$ as
\begin{equation}
\frac{\sum\limits^{}_{j\in \alpha}A_{ij}-\sum\limits^{}_{j\notin \alpha}A_{ij}}{d_i}.
\label{eq:Dani_mod_memb}
\end{equation}
Thus, the contribution becomes negative when $i$ has more neighbours outside
$\alpha$. In case of overlapping nodes belonging to multiple communities we
 also have to divide the formula above by $q_i$, corresponding to the number 
of communities of $i$. 

A further simple criterion for decent communities is to have 
a relatively large link density. Thus, the modularity of community $\alpha$ 
is given by the average of (\ref{eq:Dani_mod_memb}) over the community members,
 multiplied by the link density inside $\alpha$ as
\begin{equation}
Q_{\alpha}^{\rm ov}=\left[\frac{1}{N_{\alpha}}\sum_{i\in \alpha}\frac{\sum\limits^{}_{j\in \alpha}A_{ij}-\sum\limits^{}_{j\notin \alpha}A_{ij}}{d_i\cdot q_i}\right] \frac{M_{\alpha}}
{{N_{\alpha} \choose 2}},
\label{eq:Dani_mod_com}
\end{equation}
where $N_{\alpha}$ and $M_{\alpha}$ denote the number of nodes and links inside
$\alpha$ respectively. The overall modularity of a given partition is simply
the average of $Q_{\alpha}^{\rm ov}$ over the communities given by
\begin{equation}
Q^{\rm ov}=\frac{1}{K}\sum_{\alpha=1}^KQ_{\alpha}^{\rm ov}=\frac{1}{K}\sum_{\alpha=1}^K\frac{\sum\limits^{}_{i\in \alpha}\frac{\sum\limits^{}_{j\in \alpha}A_{ij}-\sum\limits^{}_{j\notin \alpha}A_{ij}}{d_i\cdot q_i}}{N_{\alpha}}\frac{M_{\alpha}}
{{N_{\alpha} \choose 2}}.
\label{eq:Dani_modul}
\end{equation}
Since ${1 \choose 2}$ is not defined, $Q_{\alpha}^{\rm ov}$ for communities
corresponding to single nodes is zero by definition. However, in order to
avoid partitions having only a few communities with very high $Q_{\alpha}^{\rm ov}$ values, L\'az\'ar et al. suggested collecting all unclassified nodes and 
communities corresponding to single nodes into a separate community. 
A nice feature of (\ref{eq:Dani_mod_memb}-\ref{eq:Dani_modul}) is that 
it does not require normalised membership values for the nodes, which makes
it possible that e.g., high degree members can contribute more to the 
modularity of their community compared to low degree nodes.

We note that a slight drawback of (\ref{eq:Dani_modul}) is that it does 
not take into account the size of the communities, every community is 
treated equally when calculating the average. This can cause problems
when a giant community
has emerged spreading over the whole system in the following way. The
 individual modularity for the giant community
given by (\ref{eq:Dani_mod_com}) is almost surely low, since 
the link density inside
cannot be significantly larger than the overall link density in the network.
However, as long as there are still at least a few small good quality 
communities around, the modularity of the whole system can remain
high if the contributions from the small communities 
to (\ref{eq:Dani_modul}) suppress the single
low quality contribution from the giant community. 
In order to prevent very large communities from ``hiding'' their
contribution in the overall modularity in this manner we propose an
alternative version for 
for $Q^{\rm ov}$. Instead of treating the communities equally, we weight them 
by the fraction of contained links obtaining
\begin{equation}
\widehat{Q}^{\rm ov} = \sum_{\alpha=1}^K \frac{M_{\alpha}}{M}Q_\alpha^{\rm ov}.
\label{eq:Dani_mod_mod}
\end{equation}
The weighted average above is very similar in nature 
to (\ref{eq:partition_dens}), used for calculating the 
overall partition density $D$ from the $D_{\alpha}$ defined for the individual
communities. 

\section{Applications}
\label{sect:Appl}

We studied the behaviour of the overlapping modularities described in 
Sect.\ref{sect:modul}. for partitions obtained by the CPM in a 
couple of real networks. As mentioned in Sect.\ref{sect:intro}., the
question of main interest here is whether we find a (local or even 
global) maximum in the modularity in the vicinity of the critical point
of $k$-clique percolation. Since the networks we studied were all weighted,
our method for tuning the system to the critical point was 
the application of a weight threshold, as explained in Sect.\ref{sect:CPM}.
However, due to the different origin, the total range and the distribution
of the link weights was varying from system to system. To treat all
networks we investigated in the same framework, we ordered
the links in each system according to their weights, and subsequently 
removed them in this order starting from the lowest link weights. 
For this deterministic removal process 
the control parameter of the phase transition is 
given by the fraction of removed links, $f$.
(A considerable advantage of this approach is
that it can be used also for un-weighted networks 
with random link removal
processes).

For each given value of $f$, the communities were extracted with the 
help of CFinder \cite{CFinder}, a freely downloadable implementation of 
the CPM. To monitor the $k$-clique percolation transition, we calculated 
the relative size of the largest community $\alpha_G$ given by 
$S_{G}\equiv N_{\alpha_G}/N$, corresponding 
to the order parameter, (which is 1 if the largest community
includes all nodes, and is of the order of $1/N$ when $f\rightarrow 1$.) 
However, probably the most widely used method for determining the critical fraction of removed
links $f_c$ is
via the susceptibility, $\chi$. 
This quantity can be defined as the expected change in the 
size of the largest community $\alpha_G$ when merging 
with another community chosen at random with a probability proportional 
to the community size:
\begin{equation}
\chi =\sum_{\alpha\neq\alpha_G}\frac{N_{\alpha}^2}{\sum\limits^{}_{\alpha \neq \alpha_G}N_{\alpha}}.
\end{equation}
At the critical point $\chi$ is diverging in the thermodynamic limit, 
which is manifested in a sharp peak for finite size systems.

Beside calculating $S_G$ and $\chi$, for a given value of the removed
fraction of links $f$ we also evaluated the overlapping modularity measures
discussed in Sect.\ref{sect:modul}., (taking into account only the remaining links and omitting the already removed ones). 
For clarity, in Table \ref{table:mods}. 
we summarise their notion and defining formulae used in this paper. 

\begin{table}[b!]
\begin{center}
\begin{tabular}{|l|l|}
\hline 
$Q^f$ & Fuzzy modularity by Nepusz et al. \cite{Nepusz_fuzzy} given in Eq.(\ref{eq:fuzzy_modul}), \\
  & where the partition matrix $u_{\alpha i}$ is evaluated according to Eq.(\ref{eq:part_matr_simp}) \\
\hline 
$Q^C$ & Fuzzy modularity by Chen et al. \cite{Chen_version}, given in Eq.(\ref{eq:fuzzy_modul}), \\
 & where the partition matrix $u_{\alpha i}$ is evaluated according to Eq.(\ref{eq:part_matr_Chen}) \\
\hline
$Q^S$ & Fuzzy modularity by Shen et al. \cite{Shen_modul}, given in Eq.(\ref{eq:fuzzy_modul}), \\
 & where the partition matrix $u_{\alpha i}$ is evaluated according to Eq.(\ref{eq:part_matr_Shen}) \\
\hline  
$D$ & Partition density by Ahn et al. \cite{Sune_nature}, given in Eq.(\ref{eq:partition_dens}) \\
\hline  
$Q^{\rm ov}$ & Overlapping modularity by L\'az\'ar et al. \cite{Dani_modul},
 given in Eq.(\ref{eq:Dani_modul}) \\
\hline
$\widehat{Q}^{\rm ov}$ &  Overlapping modularity by L\'az\'ar et al. modified,
 given in Eq.(\ref{eq:Dani_mod_mod})\\
\hline 
\end{tabular}
\end{center}
\caption{Summary of the examined overlapping modularity measures. 
The three fuzzy modularities, $Q^f$, $Q^C$ and $Q^S$ 
are very similar in nature, originating
from the $Q$ introduced by Girvan and Newman \cite{Newman_modul} and 
formulated in Eq.(\ref{eq:fuzzy_modul}). The only difference between them is
in the evaluation of the partition matrices. The partition density $D$ was 
introduced for ``link-based'' communities, and hence, it is based on the 
internal link density of communities. Finally, the overlapping modularity
by L\'az\'ar et al., $Q^{\rm ov}$, and also its modified version 
$\widehat{Q}^{\rm ov}$ proposed here provide a third alternative approach by 
combining different requirements for overlapping communities.}
\label{table:mods}
\end{table}
The family of fuzzy modularities, $Q^f$, $Q^C$ and $Q^S$ are generalisations
of the original non-overlapping modularity by Girvan and Newman \cite{Newman_modul}. The sole difference between these three
quantities lies in the evaluation of the partition matrices, $u_{\alpha i}$, 
needed for the calculation of the similarity $s_{ij}$ between the nodes 
in Eq.(\ref{eq:similarity}). The obtained $s_{ij}$ are then plugged 
into Eq.(\ref{eq:fuzzy_modul}) for all three quantities. The partition density,
 $D$, was introduced by Ahn et al.\cite{Sune_nature} for measuring the 
quality of overlapping ``link-communities'', and hence, it relies heavily
on the link density inside the communities. Finally, the original- and slightly
modified version of the overlapping modularity by L\'az\'ar et al.
\cite{Dani_modul} were formulated by mixing multiple requirements towards
meaningful overlapping communities.

Furthermore, we also evaluated the modularities when the giant percolating 
community was omitted in the calculation. A community spreading over the 
vast majority of the nodes is a singular object. In contrast, the 
modularity measures were designed for 
quantifying meaningful partitions, and are not expected to handle singular
communities like the giant $k$-clique percolation cluster correctly. 
A very simple idea to get around this problem is to consistently disregard
$\alpha_G$ when calculating the modularity measures.

\subsection{The studied networks}
The list of studied networks was the following:
\begin{itemize}
\item The social network between students of the University of California, 
(UNICAL), constructed from an online message record \cite{Unical_net}. 
The database contained 1899 
users forming altogether 13833 connections, where the weight of a link
corresponded to the total number of characters sent between the two endpoints. 
\item The social network between scientists based on co-authorship given
 by publications in the Los-Alamos e-print archive under ``astro-ph'' (Astrophysics), from 1995 to 1999 \cite{astro-ph}. Here the network was obtained   
by projecting the bipartite graph of authors and articles onto 
the single mode graph of authors. The link weights were calculated 
according to
\begin{equation}
w_{ij}=\sum_{k}\frac{\widehat{A}_{ik}\widehat{A}_{jk}}{\sum_l\widehat{A}_{lk}-1},
\end{equation}
where $\widehat{A}_{ik}$ denotes the adjacency matrix of the bipartite graph,
 (i.e., $\widehat{A}_{ik}=1$ if scientist $i$ is a co-author of article $k$).
The resulting co-authorship network contained 16046 nodes and 121251 links. 
\item The word association network obtained from the
South Florida Free Association norms list (containing
 10617 nodes and 63788 links), where the weight of a link from one word to 
another indicated the frequency that the people in the survey 
associated the end point of the link with its start point 
\cite{word_assoc_net}. Since we were
interested in un-directed networks, the final weight of the links corresponded
to the sum of the weights in the two opposite directions.
\end{itemize}

\subsection{Results}

In this section we show the results obtained by omitting the giant community
in the calculation of the fuzzy modularities. 
The same figures for the ``full'' modularities including also $\alpha_G$ 
are given in the Appendix. We present our findings only for
$k=3$ or $k=4$, since the low number and small size of the communities 
at larger $k$ values made the precise location of the critical point 
impossible in the systems we studied. In all of our experiments, the resolution in the removed fraction of links, $f$, was set to $0.005$.

We begin with the results for the word association network 
in Fig.\ref{fig:word_assoc}., since the critical behaviour of the $k$-clique
percolation transition is a far more transparent and articulate here
compared to the other systems investigated in this paper.

\afterpage{
\begin{figure}
\centerline{\includegraphics[width=0.7\textwidth]{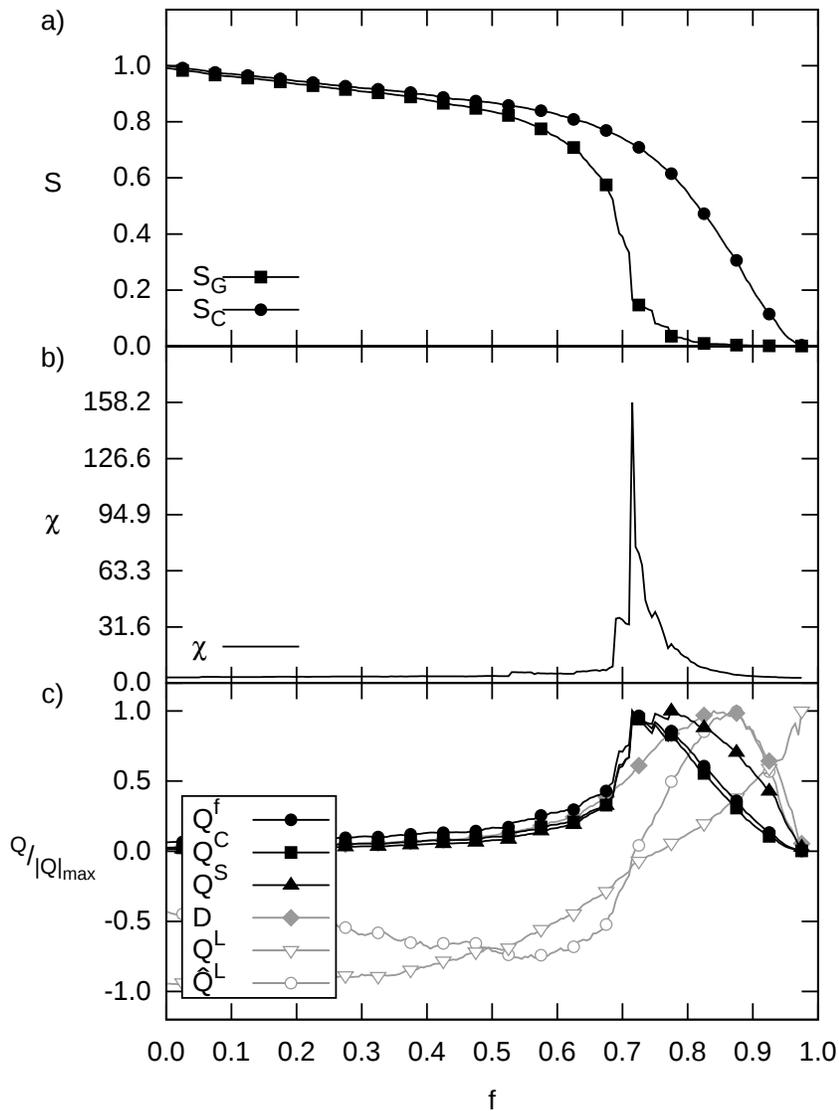}}
\caption{Results for the word association network at $k=3$, 
with a resolution in the fraction of removed links, $f$, set to $0.005$. (Please note that symbols on the plots appear at a smaller frequency, as they are intended only for making the different curves more distinguishable).
a) The relative size 
of the largest community, $S_G$, and the relative size of the total coverage
of the communities, $S_C$, as functions of $f$.
  b) The susceptibility, $\chi$, as a function 
of $f$, showing a significant sharp peak at the critical point. 
c) The different overlapping modularities, each scaled by its maximal value,
 as functions of $f$. 
The $Q^f$, $Q^C$ and $Q^S$ have maximums very close to the critical point, and 
these maximums are quite very pronounced as well. The partition density $D$ and 
the modified modularity by L\'az\'ar et al., $\widehat{Q}^{\rm ov}$, have 
also significant maximums close to the critical point, however their position
is slightly shifted towards higher $f$. The original modularity by 
L\'az\'ar et al., $Q^{\rm ov}$ shows a more or less monotonously increasing 
tendency as a function of $f$.}
\label{fig:word_assoc}
\end{figure}
\clearpage
}

In Fig.\ref{fig:word_assoc}a we plot
the relative size of the largest community, $S_G$, (corresponding to
the order parameter of the $k$-clique percolation transition), as a function
of  $f$, at $k=3$. 
In this 
panel we also show the relative size of the total coverage of communities
in the network, denoted by $S_C$. (Due to the community definition originating
in $k$-cliques, a part of the nodes may not belong to any communities in
case of the CPM, and the size of this fraction is given by $1-S_C$.) 
Starting with a value very close to $1$ at $f=0$, both $S_G$ and $S_C$ show 
a slowly decreasing tendency, turning into a steep decay in the vicinity
of the critical point, where the two curves separate from each other
due to the faster change in $S_G$. To pinpoint the critical point 
of the $k$-clique percolation more precisely, in 
Fig.\ref{fig:word_assoc}b we display the susceptibility $\chi$ as a function
of $f$, showing a very sharp peak at $f_c$. 
The behaviour of the various overlapping modularities 
can be followed in Fig.\ref{fig:word_assoc}c,
where each modularity measure is rescaled by its maximal value and is plotted
as a function of $f$. The fuzzy modularities $Q^f$, $Q^C$ and $Q^S$ show 
rather smooth curves with single (global) maximums very close to $f_c$. 
Although we can observe an unimodal shape for also the partition 
density $D$, in this case the position of the maximum is shifted slightly
towards higher $f$ values. Nevertheless, this maximum is also consistent with
$f_c$. Interestingly, the modularity by L\'az\'ar et al., $Q^{\rm ov}$ shows a
monotonously increasing tendency as a function of $f$, with no maximum 
in the vicinity of the critical point. In our opinion, this is due to
the equal treatment of the communities irrespectively of the community size,
``hiding'' the giant community with low quality among
the better quality communities of normal size. When switching from $Q^{\rm ov}$
to $\widehat{Q}^{\rm ov}$, also taking into account the community sizes in
the averaging, we regain the unimodal shape with a maximum very close to the 
maximum of $D$, in consistency with the position of the critical point.

\afterpage{
\begin{figure}
\centerline{\includegraphics[width=0.7\textwidth]{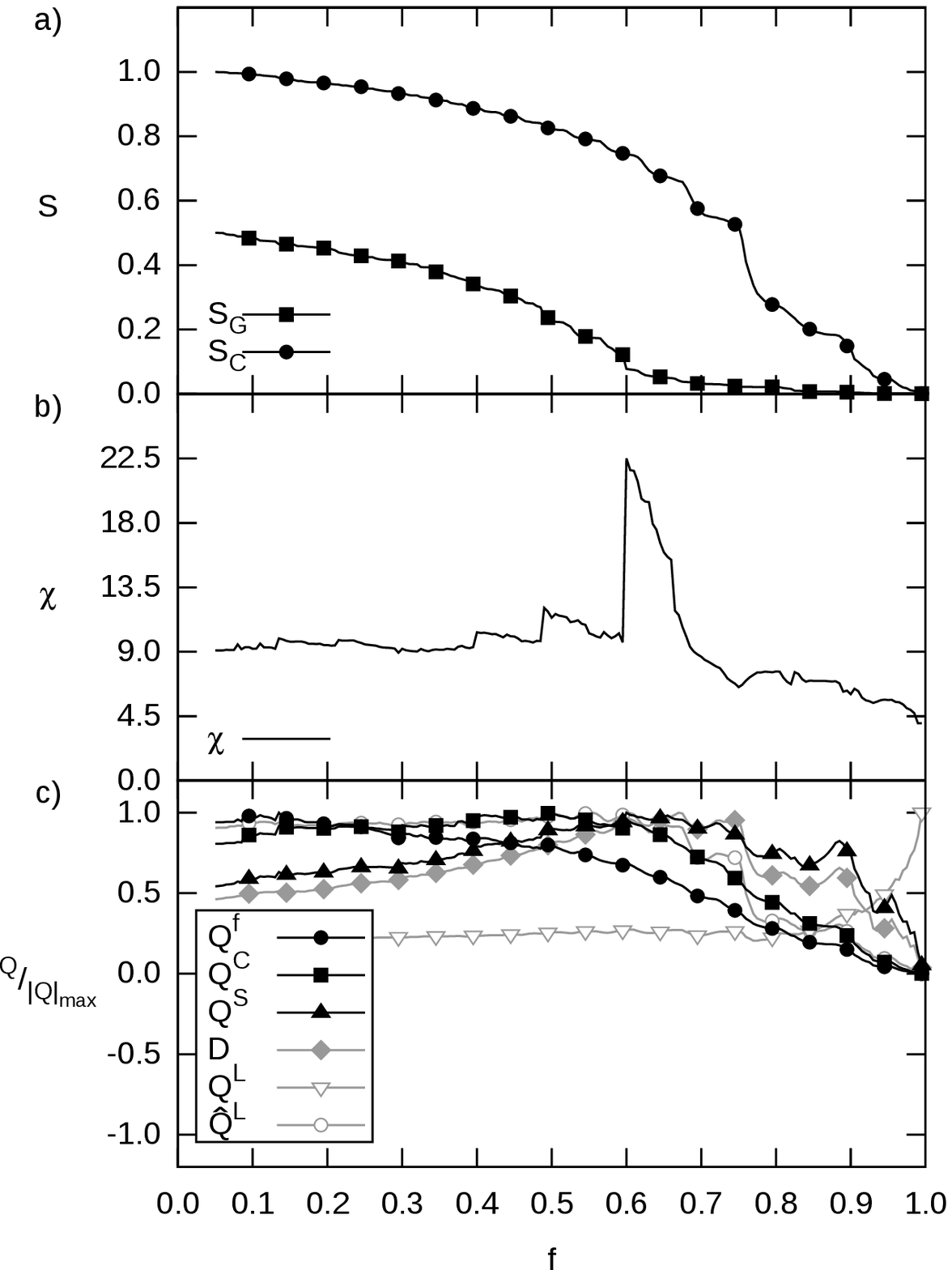}}
\caption{Results for the co-authorship network at $k=4$, with a resolution in the fraction of removed links, $f$, set to $0.005$. 
a) The relative size 
of the largest community, $S_G$ and the relative size of the total coverage
of the communities, $S_C$ in the network as functions of $f$. 
Note that in this case $S_G$  remains
significantly lower than 1 in the whole range of $f$. 
b) The susceptibility, $\chi$, as a function 
of $f$. The peak signalling the critical point is far less significant
compared to the case of the word association network shown in Fig.\ref{fig:word_assoc}.
c) The different overlapping modularities, each scaled by its maximal value,
 as functions of $f$. The $Q^f$ shows a decreasing tendency in the entire 
$f$ range, while $Q^C$ has a very weak and protracted maximum in the vicinity
of $f_c$. This maximum is more prominent (and is closer to the critical 
point) for $Q^S$ and also for the partition density $D$. 
Finally, the original modularity by L\'az\'ar et al. does not show any relevant maximum, while 
in case of $\widehat{Q}^L$ we can observe a very weak maximum similar to that 
of $Q^C$.}
\label{fig:astro-ph}
\end{figure}
\clearpage
}

In Fig.\ref{fig:astro-ph}. we display our results for the co-authorship
network at $k=4$. The percolation transition here is far less pronounced 
compared to the word association network. E.g., the relative size
of the giant community, $S_G$ is far below $1$ even at $f=0$, as shown in
Fig.\ref{fig:astro-ph}a. Although we observe a peak in the susceptibility,
$\chi$, (Fig.\ref{fig:astro-ph}b), its width compared to 
its magnitude reveals a significantly broader nature compared to the 
very sharp peak seen in Fig.\ref{fig:word_assoc}b. The corresponding 
overlapping modularities (scaled by their maximal values) are depicted
in Fig.\ref{fig:astro-ph}c, as functions of $f$. Interestingly,  
$Q^S$ and the partition density, $D$ have global maximums consistent with
the critical $f_c$, and also a very weak global maximum can be observed 
slightly below $f_c$ for $Q^C$ and $\widehat{Q}^L$. 
In contrast, the rest of the modularities show no relevant
maximum. (Extrema at either $f=0$ or $f=1$ are discarded). 

\afterpage{
\begin{figure}[hbt]
\centerline{\includegraphics[width=0.7\textwidth]{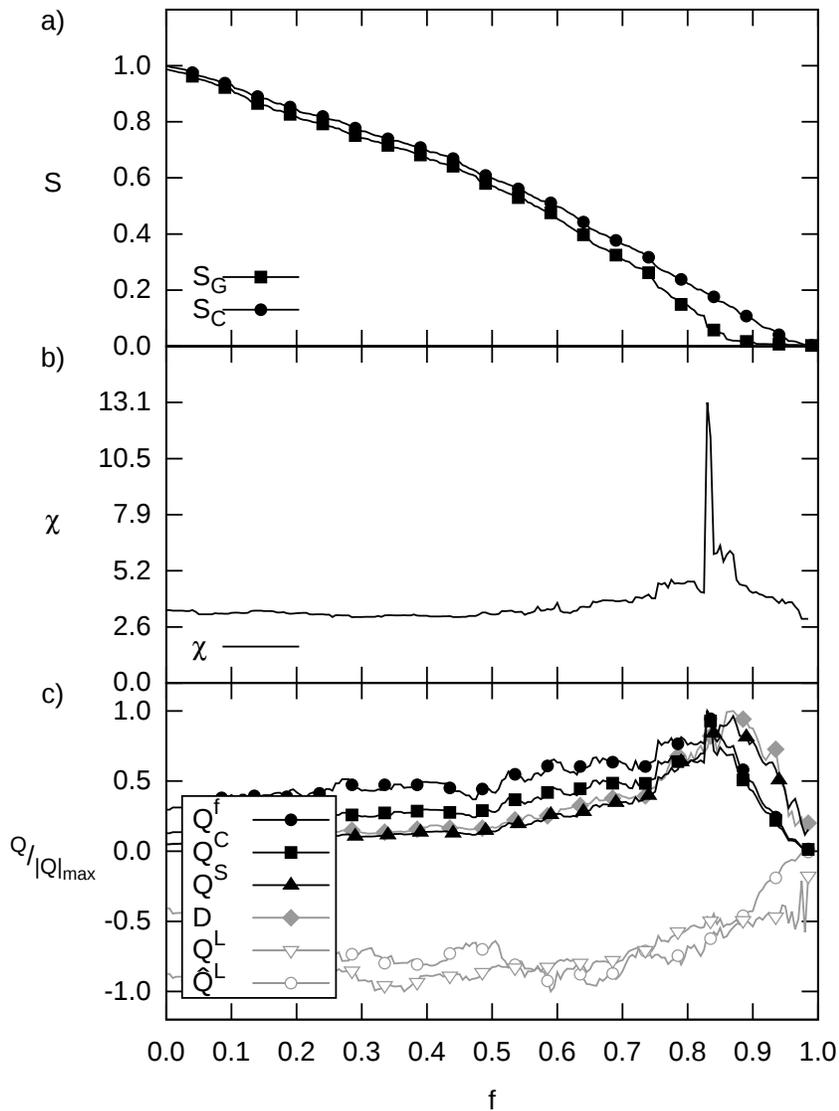}}
\caption{Results for the UNICAL network at $k=3$, with a resolution in the fraction of removed links, $f$, set to $0.005$. 
a) The relative size 
of the largest community, $S_G$ and the relative size of the total coverage
of the communities, $S_C$ in the network as functions of $f$. 
b) The susceptibility, $\chi$, as a function 
of $f$. Similarly to the co-authorship network, the peak at the critical
 point is far less pronounced here compared to word-association network
shown in Fig.\ref{fig:word_assoc}.
c) The different overlapping modularities, each scaled by its maximal value,
 as functions of $f$. In this case the fuzzy modularities $Q^f$, $Q^C$, 
$Q^S$, and also the partition density have a roughly unimodal shape with
a global maximum very close to the critical point. The $Q^{\rm ov}$ and 
$\widehat{Q}^{\rm ov}$ have rather varying shapes without any relevant 
global maximum.}
\label{fig:unical}
\end{figure}
\clearpage
}

Finally, in Fig.\ref{fig:unical}. we show the results for the UNICAL network.
Similarly to the co-authorship network, the $k$-clique percolation transition
is far less pronounced compared to the word-association network. Although
the relative size of the giant community, $S_G$ is reaching 1 at $f=0$, 
(Fig\ref{fig:unical}a), the overall shape of the curve shows a constant decay
between $f=0$ and $f_c$, in contrast to the very slow decay turning
into a sudden drop at the critical point seen in 
Fig.\ref{fig:word_assoc}a. Furthermore, the susceptibility, $\chi$
displayed in Fig.\ref{fig:unical}b shows a peak with a height smaller by 
an order of magnitude acompanied by roughly the same width 
compared to the peak in Fig.\ref{fig:word_assoc}b 
for the word association network. In Fig.\ref{fig:unical}c we plotted the 
corresponding overlapping modularities as functions of $f$. All three 
fuzzy modularities ($Q^f$, $Q^C$ and $Q^S$) and also the partition density
$D$ show slightly varying curves with an overall unimodal shape, and their
maximums is very close to the critical point. In contrast, $Q^{\rm ov}$ and $\widehat{Q}^{\rm ov}$ display no relevant maximum.

According to our results detailed in the Appendix, when including also
the giant community in the evaluation of the modularities, the maximum
for most of the measures is shifted either to $f=0$ or to $f=1$. 
These correspond to trivial optima, i.e., either all of the links have 
to be kept, or all of them has to be deleted to achieve maximal modularity.
An interesting exception is provided by the partition density, $D$, for which
we observed maximums in full consistency with the critical point for
all networks we investigated. 

In summary, the overall behaviour of the various different 
overlapping modularities
support the basic assumption that the optimal partitioning for the CPM is
obtained in the vicinity of the critical point of $k$-clique percolation. 
However, the consistency between the position of the critical point and
the maximums of the modularities can be best observed for systems 
showing a sharp, fully fledged phase transition. Furthermore, one has 
to take into account how the modularities are actually evaluated. 
The most reassuring results were given by the partition density, $D$, 
providing a global maximum always in consistency with the critical point. 
When omitting the giant percolating community from the calculation,
the fuzzy modularities, $Q^f$, $Q^C$ and $Q^S$ also showed a prominent 
maximums very close to $f_c$ in case of the word association network 
and the UNICAL network, while $Q^C$ and $Q^S$ showed consistency with
the critical point also for the co-authorship network. 
The original overlapping modularity by L\'az\'ar et al.,
 $Q^{\rm ov}$ showed a rather monotonous behaviour in all systems we 
investigated, 
lacking any relevant maximum other than $f=0$ or $f=1$. This is mainly due
to the fact that the averaging over the community-vise individual modularities 
in (\ref{eq:Dani_modul}) does not take into account the community size.
When switching to $\widehat{Q}^{\rm ov}$ incorporating size dependent 
weights, we observed a significant maximum close to the critical point 
in case of the word association network, and this maximum remained 
at its place even when including also the giant community 
in the calculation.

\section{Conclusions}
\label{sect:Concl}
Motivated by the importance of overlapping community finding 
methods in social networks, we studied the behaviour of various 
overlapping modularity measures in the vicinity of the 
critical point of $k$-clique percolation. According to our analysis of
real social- and other networks, the overlapping modularities showed
large maximums close to the critical point when the 
critical behaviour of the phase transition was prominent and articulate.
However in some of the networks we could observe only blurred, slightly
ambiguous phase transitions. Nevertheless, a part of the involved
modularity measures still displayed a maximum in consistency 
with the likely position of the critical point, while the others showed
no relevant maximums at all. 
These findings provide a strong quantitative validation for the
former heuristic for setting the parameters of the CPM, 
suggesting that the quality of the partitioning is best in the 
vicinity of the critical point of $k$-clique percolation.

\begin{emph}
This work was supported by the European Union and co-financed by the
European Social Fund (grant agreement no. TAMOP 4.2.1/B-09/1/KMR-2010-0003) 
and by the Hungarian National Science Fund (OTKA K105447).
\end{emph}

\clearpage
\section*{Appendix}

Here we show the results for the modularities presented in 
Figs.\ref{fig:word_assoc}-\ref{fig:unical}. when the giant 
percolating community is also included in the calculation.
For simplicity we also re-plotted the relative size of 
the giant component, $S_G$, and the susceptibility, $\chi$,
as functions of $f$ for each system. 

In Fig.\ref{fig:word_assoc_wgc}. we show the results for the word association
network. The three fuzzy modularities, ($Q^f$, $Q^C$ and $Q^S$), show 
an overall decreasing tendency with a strong change in the slope at
the critical point for $Q^C$ and $Q^S$. In contrast, the partition 
density $D$ and the two variations for the overlapping modularity by
L\'az\'ar et al., $Q^L$ and $\widehat{Q}^L$ behave similarly to the 
case shown in Fig.\ref{fig:word_assoc}c: $D$ and $\widehat{Q}^L$ have
prominent maximums in consistency with $f_c$, while $Q^L$ shows an increasing
tendency as a function of $f$.

\afterpage{
\begin{figure}[hbt] 
\centerline{\includegraphics[width=0.7\textwidth]{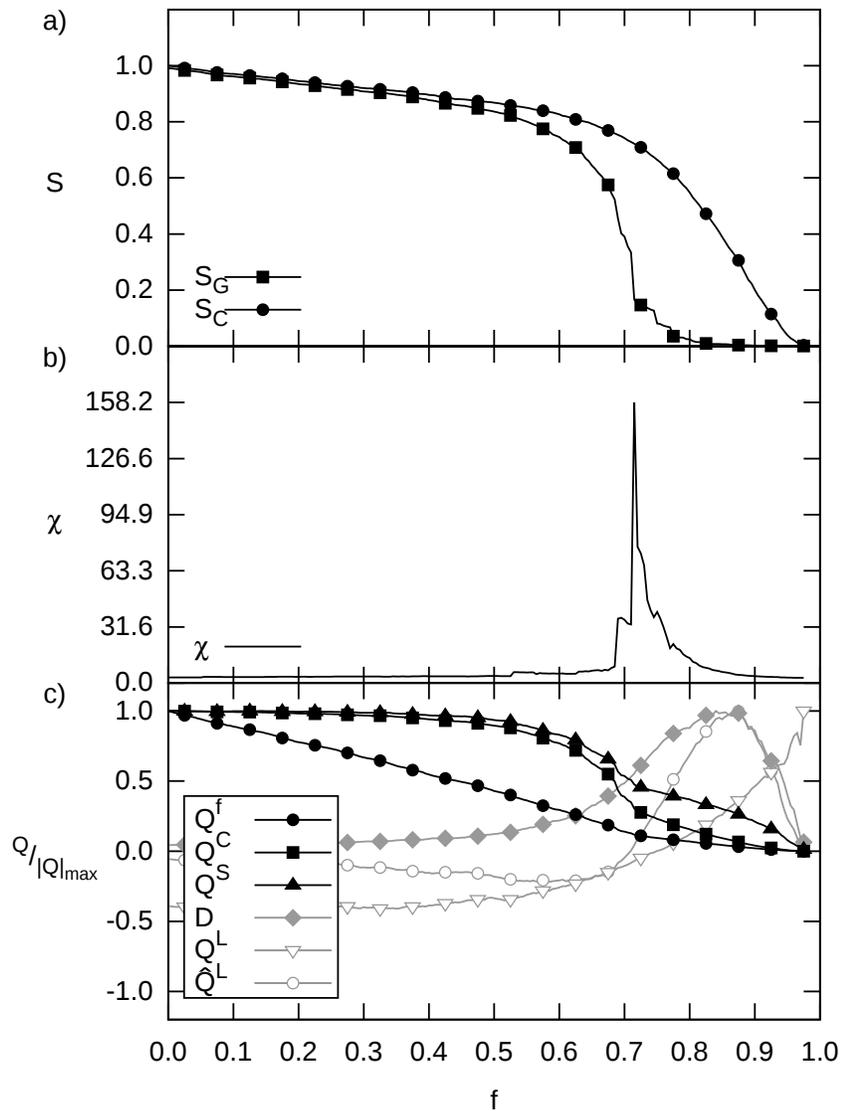}}
\caption{Results for the word association network at $k=3$ when including
also $\alpha_G$ in the evaluation of the modularities. 
a) The relative size and relative coverage of $\alpha_G$, as functions 
of the fraction
of removed links, $f$, (same as in Fig.\ref{fig:word_assoc}a). 
b) The susceptibility, $\chi$, as a function 
of $f$ (same as in Fig.\ref{fig:word_assoc}b). 
c) The different overlapping modularities, each scaled by its maximal value,
 as functions of $f$. 
The $Q^f$, $Q^C$ and $Q^S$ show a decreasing tendency, with a more steep
drop at $f_c$ in case of $Q^C$ and $Q^S$. The partition density $D$ and 
the modified modularity by L\'az\'ar et al., $\widehat{Q}^{\rm ov}$, still have 
 significant maximums at a position consistent with the critical point. 
The original modularity by 
L\'az\'ar et al., $Q^{\rm ov}$ shows an increasing 
tendency as a function of $f$, similarly as in case of 
Fig.\ref{fig:word_assoc}c.}
\label{fig:word_assoc_wgc}
\end{figure}
\clearpage
}

In Fig.\ref{fig:astro-ph_wgc}. we display the results for
the co-authorship network. Here the fuzzy modularities, 
($Q^f$, $Q^C$ and $Q^S$), show a smoothly and slowly decreasing curve 
in the entire $f$ range. Similarly to the case of the word association
network, the shape of the curves for $D$, $Q^L$ and $\widehat{Q}^L$ are
less affected by the inclusion of $\alpha_G$ in the calculation: 
$D$ has a global maximum in the vicinity of $f_c$ and $Q^L$ shows an almost
 constant behaviour with a sudden increase for large $f$.  
For $\widehat{Q}^L$, the very weak maximum in Fig.\ref{fig:astro-ph}c has 
been flattened, however, a steeply decreasing function can still be
observed above $f_c$.

\afterpage{
\begin{figure}[hbt]
\centerline{\includegraphics[width=0.7\textwidth]{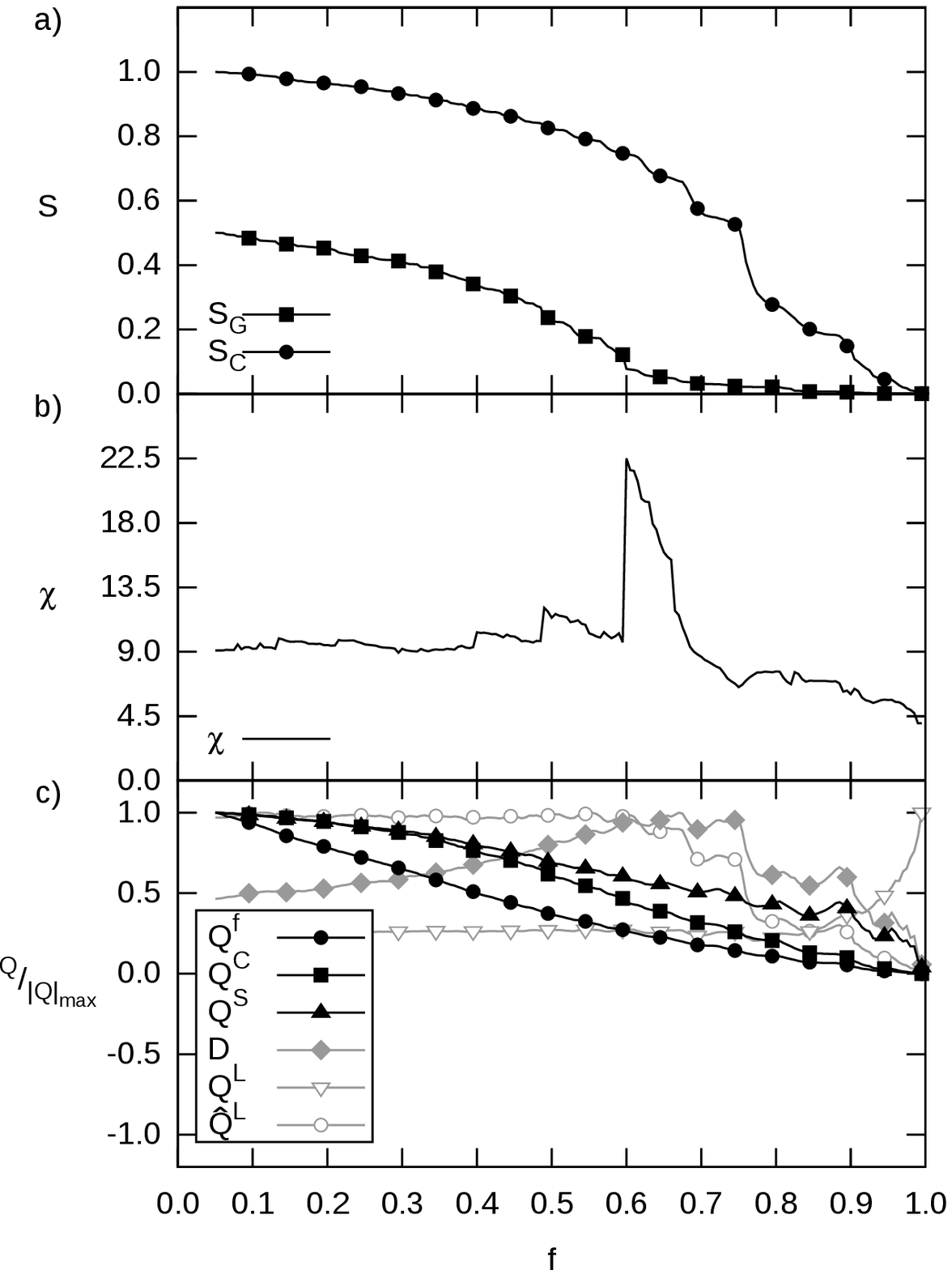}}
\caption{Results for the co-authorship network at $k=4$. 
a) The relative size and relative coverage of $\alpha_G$, as functions 
of the fraction
of removed links, $f$, (same as in Fig.\ref{fig:astro-ph}a). 
b) The susceptibility, $\chi$, as a function 
of $f$ (same as in Fig.\ref{fig:astro-ph}b)
c) The different overlapping modularities, each scaled by its maximal value,
 as functions of $f$. The fuzzy modularities, $Q^f$, $Q^C$ and $Q^S$ show a 
slowly decreasing tendency, while the partition density $D$ has a global
maximum relatively close to the critical point. The modularity $Q^L$ 
is more or less constant except for the very large $f$ region where it
becomes increasing. In contrast, $\widehat{Q}^L$ is close to $1$ below $f_c$, 
and starts decreasing above.}
\label{fig:astro-ph_wgc}
\end{figure}
\clearpage
}

Finally, in Fig.\ref{fig:unical_wgc}. we show the results for
the UNICAL network. Again, the fuzzy modularities, ($Q^f$, $Q^C$ and $Q^S$),
display a smoothly decreasing tendency. However, for $Q^C$ and $Q^S$ 
the slope becomes higher close to the critical point, similarly to
the case shown in Fig.\ref{fig:word_assoc_wgc}c. The partition density,
 $D$, has an overall unimodal shape with smaller fluctuations and a
 global maximum quite close to the critical point. The $Q^{\rm ov}$ and $\widehat{Q}^{\rm ov}$ display no relevant maximum, 
with $\widehat{Q}^{\rm ov}$ actually arriving to a minimum close to
$f_c$. The explanation of this effect is yet unresolved and waits
for future work.

\afterpage{
\begin{figure}[hbt]
\centerline{\includegraphics[width=0.7\textwidth]{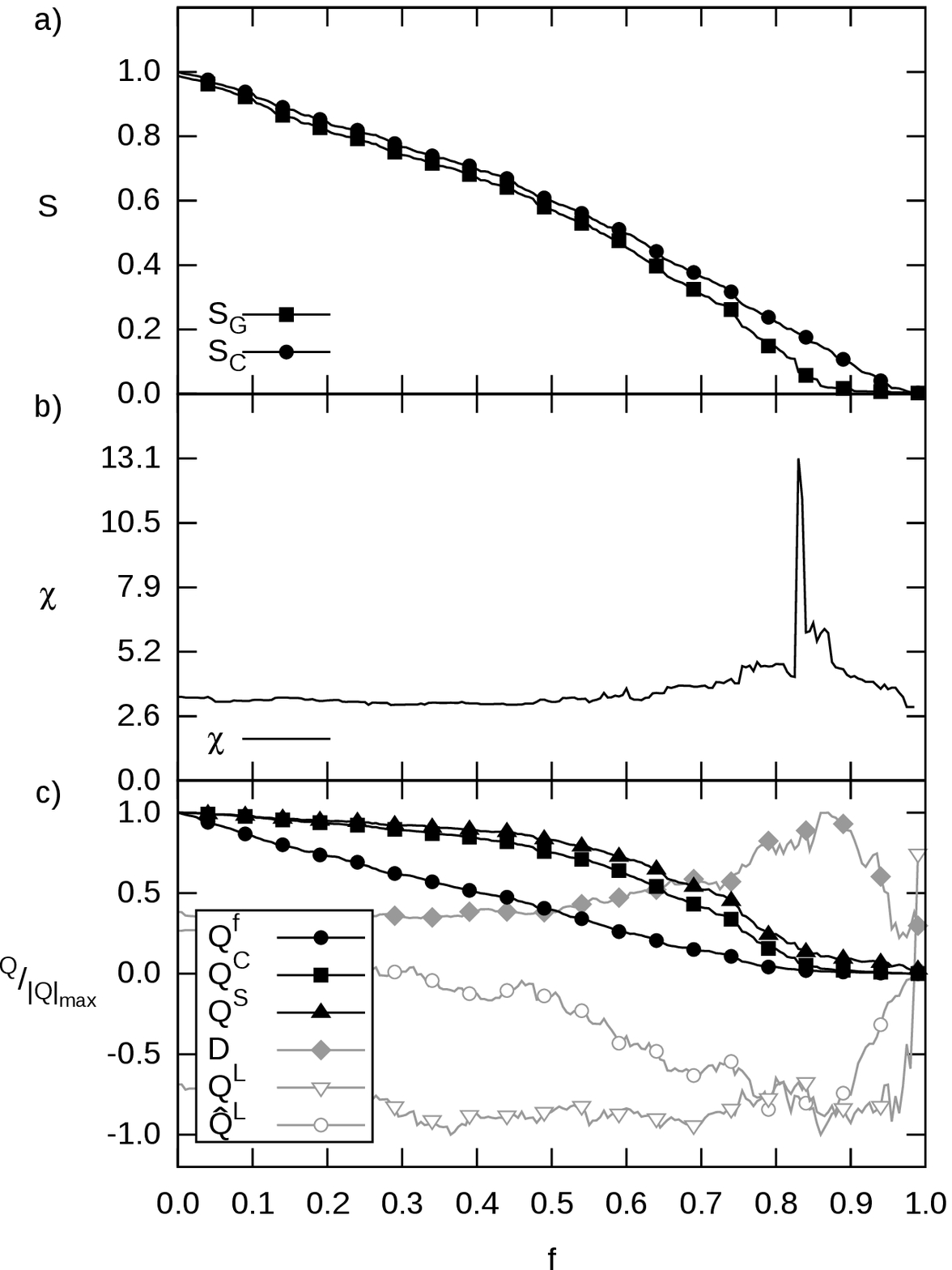}}
\caption{Results for the UNICAL network at $k=3$. 
a) The relative size and relative coverage of $\alpha_G$, as functions 
of the fraction
of removed links, $f$, (same as in Fig.\ref{fig:unical}a). 
b) The susceptibility, $\chi$, as a function 
of $f$ (same as in Fig.\ref{fig:unical}b).
c) The different overlapping modularities, each scaled by its maximal value,
 as functions of $f$. Similarly to Fig.\ref{fig:word_assoc_wgc}c, the
fuzzy modularities $Q^f$, $Q^C$ and $Q^S$ show a decreasing tendency with
a steeper drop in $Q^C$ and $Q^S$ at the critical point. The partition
density, $D$ has retained its overall unimodal shape with a global maximum
in consistency with $f_c$. The original- and modified modularity 
by L\'az\'ar et al. show rather varying curves with no relevant maximum.
}
\label{fig:unical_wgc}
\end{figure}
\clearpage
}

\clearpage

\bibliographystyle{unsrt}
\bibliography{ov_mod_sp_arxiv}

\end{document}